\documentclass[showpacs,prl,twocolumn,aps,superscriptaddress,preprintnumbers,letterpaper]{revtex4}
\usepackage{amsmath,amssymb,bm}
\usepackage{epsfig}
\usepackage{graphicx}
\usepackage{amsfonts}
\usepackage{epstopdf}
\def\slashchar#1{\setbox0=\hbox{$#1$}     		
   \dimen0=\wd0                                 	
   \setbox1=\hbox{/} \dimen1=\wd1               	
   \ifdim\dimen0>\dimen1                        	
      \rlap{\hbox to \dimen0{\hfil/\hfil}}      	
      #1                                        	
   \else                                        	
      \rlap{\hbox to \dimen1{\hfil$#1$\hfil}}   	
      /                                         	
   \fi}

\renewcommand{\vec}{\boldsymbol}
\newcommand{\be}{\begin{equation}}
\newcommand{\ee}{\end{equation}}
\newcommand{\bear}{\begin{eqnarray}}
\newcommand{\eear}{\end{eqnarray}}
\newcommand{\ba}{\begin{array}}
\newcommand{\ea}{\end{array}}

\renewcommand\a{\alpha}

\renewcommand\d{\delta}

\renewcommand\r{\rho}

\renewcommand\c{\chi}

\renewcommand\o{\omega}

\newcommand\g{\gamma}

\newcommand\m{\mu}

\newcommand\p{\pi}

\newcommand\s{\sigma}

\renewcommand\L{\Lambda}

\renewcommand\S{\Sigma}

\newcommand\G{\Gamma}

\newcommand{\fig}[1]{Fig.~\ref{#1}}
\newcommand{\eq}[1]{Eq.~(\ref{#1})}

\newcommand\lb{\left(}
\newcommand\rb{\right)}
\newcommand\ls{\left[}
\newcommand\rs{\right]}

\newcommand{\lan}{\langle}
\newcommand{\ran}{\rangle}

\newcommand\ra{\rightarrow}

\newcommand{\non}{\nonumber\\}

\newcommand{\Tr}{{\rm Tr}}

\newcommand{\im}{{\rm{Im}}}

\newcommand{\bp}{{\vec p}}
\newcommand{\bk}{{\vec k}}

\renewcommand{\part}{{\rm part}}

\renewcommand{\vec}{\boldsymbol}

\begin{document}

\title{Axial Current Generation from Electric Field: Chiral Electric Separation Effect}

\author{Xu-Guang Huang}
\email{xuhuang@indiana.edu}
\affiliation{ Physics Department and Center for Exploration of Energy and Matter,
Indiana University, 2401 N Milo B. Sampson Lane, Bloomington, IN 47408, USA.}

\author{Jinfeng Liao}
\email{liaoji@indiana.edu}
\affiliation{ Physics Department and Center for Exploration of Energy and Matter,
Indiana University, 2401 N Milo B. Sampson Lane, Bloomington, IN 47408, USA.}
\affiliation{RIKEN BNL Research Center, Bldg. 510A, Brookhaven National Laboratory, Upton, NY 11973, USA.}

\date{\today}

\begin{abstract}
We study a relativistic plasma containing charged chiral fermions in an external electric field.
We show that with the presence of both vector and axial charge densities, the electric field can induce an axial current along its direction and thus
cause chirality separation. We call it the chiral electric separation effect (CESE). On a very general basis,
we argue that the strength of CESE is proportional to $\m_V\m_A$ with $\m_V$ and $\m_A$ the chemical potentials for
vector charge and axial charge. We then explicitly calculate this CESE conductivity coefficient in thermal QED at
leading-log order. The CESE can manifest a new gapless wave mode propagating along the electric field.
Potential observable effects of CESE in heavy-ion collisions are also discussed.
\end{abstract}
\maketitle

{\em Introduction.---}It was discovered and understood a long time ago that  an external electric field, when applied to any
conducting matter, induces a vector current, as described by Ohm's law
\be
\vec j_V= \sigma \vec E, \label{con}
\ee
where $\sigma$ is the electric conductivity of the matter, and we use the convention that the electric current is $e \vec j_V$. In quantum field theories such as the quantum electrodynamics (QED) and
the quantum chromodynamics (QCD), one has not only the vector current $j^\mu_V$ but also the axial current $j^\mu_A$ when there are charged chiral fermions. A very interesting question is then, in addition to the above conducting vector current under applied electric field, what are the other possible current generations in response to externally applied Maxwell electric and/or magnetic fields.

Recently, the QCD axial anomaly has been found to induce the following two phenomena in the high-temperature deconfined phase of QCD, the quark-gluon plasma (QGP), with the presence of an external magnetic field:
the chiral magnetic effect (CME) and the chiral separation effect (CSE).
The CME is the generation of vector current and thus the electric charge separation along the axis of the applied magnetic field in the presence of nonzero axial charge density arising from fluctuating topological charge \cite{Kharzeev:2004ey,Kharzeev:2007tn,Kharzeev:2007jp,Fukushima:2008xe,Kharzeev:2009fn}. With an imbalance
between the densities of left- and right-handed quarks, parametrized by an axial
chemical potential $\mu_A$, an external magnetic field induces the vector current $j^i_V = \lan\bar{\psi} \gamma^i \psi\ran$,
\be
\vec j_V=\sigma_5 \mu_A \vec B, \label{cme}
\ee
with chiral conductivity $ \sigma_5\equiv {N_ce \over 2\pi^2}$. A ``complementary'' effect also arising from the axial anomaly is the CSE which predicts the generation of an axial current, $j_A^i=\lan\bar{\psi} \gamma^i\gamma_5 \psi\ran$, and thus separation of axial charges along the  external $\vec B$ field at nonzero vector charge density (parametrized by its chemical potential $\mu_V$) \cite{son:2004tq,Metlitski:2005pr},
\be
\vec j_A=\sigma_5  \mu_V \vec B . \label{cse}
\ee
It should be emphasized, though, the $\mu_A$ (unlike $\mu_V$) is not associated with any conserved charge and can only be treated as an external parameter arising from external dynamics in the slowly varying limit, e.g., via effective axion dynamics $\mu_A\sim \partial\theta/\partial t \ll \L_{QCD}$; see detailed discussions in~\cite{Kharzeev:2007tn,Zhitnitsky:2012im}. This approach is justified for the study in the present Letter as also in previous studies~\cite{Kharzeev:2004ey,Kharzeev:2007tn,Kharzeev:2007jp,Fukushima:2008xe,Kharzeev:2009fn,son:2004tq,Metlitski:2005pr,Zhitnitsky:2012im}.

There is robust evidence for both CME and CSE from kinetic theory, hydrodynamics, and holographic QCD models in strong coupling regime as well as in lattice QCD computations \cite{Yee:2009vw,Rebhan:2009vc,Rubakov:2010qi,Gynther:2010ed,Gorsky:2010xu,Sadofyev:2010pr,Kalaydzhyan:2011vx,nair,Gao:2012ix,Son:2012wh,Stephanov:2012ki,Buividovich:2009wi,Abramczyk:2009gb}. Experimentally, the hot QGP is created in high-energy heavy ion collisions at the Relativistic Heavy Ion Collider (RHIC) and the Large Hadron Collider (LHC). In such collisions,domains of QGP with nonzero chirality ($\mu_A\ne 0$) can arise from topological transitions in QCD, and there are also extremely strong transient $\vec E$ and $\vec B$ fields \cite{Bzdak:2011yy,Deng:2012pc,Bloczynski:2012en,Tuchin:2010vs,Voronyuk:2011jd}, so the CME and CSE effects can occur. There have been measurements of charge asymmetry fluctuations motivated by CME predictions from the STAR \cite{:2009uh} and PHENIX \cite{Ajitanand:2010rc}
Collaborations at RHIC, as well as from the ALICE \cite{Abelev:2012pa} at LHC. The precise meaning of these data is under investigations (see, e.g., \cite{Bzdak:2009fc}). It has also been proposed that the combination of the CME and CSE leads to a collective excitation in QGP called chiral magnetic wave (CMW) \cite{Kharzeev:2010gd}. The CMW induces an electric quadruple of QGP that can be measured via elliptic flow splitting between $\pi^-/\pi^+$ \cite{Burnier:2011bf}, with supportive evidence from STAR measurements \cite{Wang:2012qs}.

There is, however, one more possibility that has not been previously discussed, namely the generation of an axial current in the electric field. We find this to occur when the matter has both nonzero $\mu_V$ and  nonzero $\mu_A$,
\be
\vec j_A = \chi_{e} \mu_V \mu_A \vec E , \label{cese}
\ee
which can be called a chiral electric separation effect (CESE). In this Letter, we will derive this relation, and explicitly compute the CESE conductivity $\chi_{e}$ in thermal QED. With this new relation found, one can nicely combine all four effects into the following form:
\be
\left(\begin{array}{c} \vec j_V \\ \vec j_A\end{array}\right) =  \left(\begin{array}{cc}
\sigma  & \sigma_5 \mu_A \\   \chi_e \mu_V \mu_A &  \sigma_5 \mu_V
\end{array}\right) \left(\begin{array}{c}
\vec E \\ \vec B
\end{array}\right). \label{vaeb}
\ee

{\em Chiral electric separation effect.---} To intuitively understand how the CESE (\ref{cese}) arises, let us consider a conducting system with chiral fermions. When an electric field is applied, the positively or negatively charged fermions will move parallel or antiparallel to the $\vec E$ direction and both contribute to the total vector current as in Eq.(\ref{con}). If $\mu_V>0$ then there will be more positive fermions (moving along $\vec E$), and further if $\mu_A>0$, then there will be more right-handed fermions than left-handed ones. The end result will thus be a net flux of right-handed (positive) fermions moving parallel to $\vec E$. This picture is most transparent in the extreme situation when the system contains only right-handed fermions (i.e., in the limit of $\mu_V=\mu_A>0$), with both a vector and an axial current concurrently generated parallel to $\vec E$. The same conclusion can be made when both $\mu_V$ and $\mu_A$ are negative.
  In cases with $\mu_V>0>\mu_A$ or $\mu_V<0<\mu_A$ one can follow the same argument to see an axial current generated antiparallel to the $\vec E$ direction.

Different from the CME and CSE, the generation of the axial current via CESE is not related to axial anomaly, but rather arises from the same conducting transport responsible for usual conduction in Eq.(\ref{con}). To demonstrate that, let us consider the conduction of left-handed fermions, $\vec j_L=\sigma(T;\m_L,\mu_R)\vec E$ where $\mu_L=\mu_V - \mu_A$ and $\mu_R=\mu_V+\mu_A$ are the left-handed and right-handed chemical potentials, respectively. Due to symmetry, the conducting
current of right-handed fermions must be $\vec j_R=\sigma(T;\m_R,\mu_L)\vec E$.
We focus on the situation when all chemical potentials are small compared to the temperature $T$, so that $\sigma(T;\mu_L,\mu_R)\approx \sigma_0(T) + \sigma'_0(T) \mu_L^2+\sigma''_0(T) (\mu_L^2+\mu_R^2)$ where the second term represents the chemical potential correction to the fermion conduction (here, the left-handed ones) and the last term arises from correction to the screenings (as will be seen in our explicit calculation later). Note that due to charge conjugation symmetry there will be no linear terms of $\mu_{L,R}$ in the expansion. Now the total vector current would then be $\vec j_V=\vec j_L + \vec j_R =\sigma(T;\mu_L,\mu_R) \vec E + \sigma(T;\mu_R,\mu_L) \vec E  \approx [2\sigma_0 + (\sigma'_0+2\sigma_0'') (\mu_V^2+\mu_A^2)] \, \vec E $
while the axial current would then be $\vec j_A = \vec j_R - \vec j_L \approx 4 \sigma'_0\, \mu_V \mu_A\, \vec E$. We therefore see that its origin is not from the axial anomaly but from the conduction in a chiral many-body environment, and the CESE is different in nature from recently discussed axial current generation via anomaly in certain superfluid systems \cite{Neiman:2011mj,Lin:2011mr,Newman:2005hd}.  Note also that the relevant time scale, e.g., $\sim 1/(e\sigma)$ should be long enough to justify treating $\mu_A$ as a slowly varying external parameter~\cite{Kharzeev:2007tn,Zhitnitsky:2012im}.

{\em Computation of the CESE conductivity.---}We will now compute explicitly the leading-log order CESE conductivity for thermal QED plasma using the Kubo formula under the condition $\m_V,\mu_A\ll T$. The extension of this computation to QCD is straightforward and will be presented elsewhere.

\begin{figure}[!htb]
\begin{center}
\includegraphics[width=3.5cm]{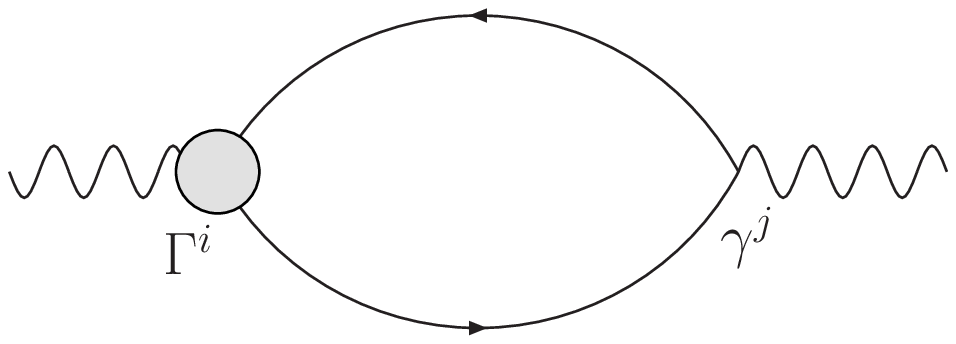}\;\;
\includegraphics[width=3.5cm]{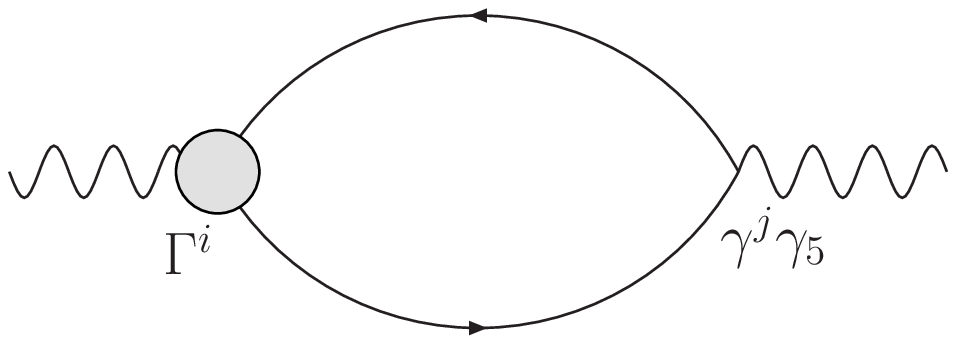}
\caption{The Feynman diagrams for retarded correlators $G_{VV}^R$ (left) and $G_{AV}^R$ (right).} \label{diag:kubo}
\vspace{-0.2in}
\end{center}
\end{figure}

\begin{figure}[!htb]
\begin{center}
\includegraphics[width=8cm]{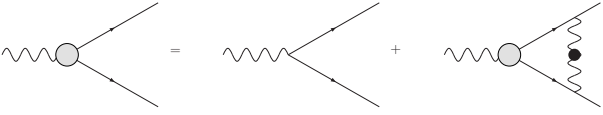}
\caption{Integral equation for the effective vertex $\G^\m$.} \label{diag:vertex}
\vspace{-0.2in}
\end{center}
\end{figure}

Let us denote $\s_e=\c_e\m_V\m_A$. Starting with  the retarded vector-vector  and vector-axial correlators $G_{VV}^{Rij}$ and $G_{AV}^{Rij}$
(as diagrammatically shown in \fig{diag:kubo}), the $\s$ and $\s_e$ are given via the Kubo formulas as
\begin{eqnarray}
\s&=&\sum_{i=1}^3\lim_{\o\ra 0}\lim_{\bk\ra0}\frac{i}{3\o}G_{VV}^{Rii}(\o,\bk),\\
\s_e&=&\sum_{i=1}^3\lim_{\o\ra 0}\lim_{\bk\ra0}\frac{i}{3\o}G_{AV}^{Rii}(\o,\bk),
\end{eqnarray}
In \fig{diag:kubo}, the shaded circle represents an effective vertex 
(see \fig{diag:vertex}). This effective vertex represents a resummation of a set of ladder diagrams which contribute to the same order owing to the pinching singularity when $\o,\bk\ra0$~\cite{Jeon:1994if}. The dotted line in \fig{diag:vertex} represents the hard thermal loop (HTL) resummed propagator. All other kinds of diagrams (e.g., the box diagrams which
are vanishing identically due to Furry theorem at zero chemical potentials but finite at nonzero chemical potentials) are sub-leading-log order.

One can write down the retarded correlators explicitly,
\begin{eqnarray}
G_{VV}^{ij}&=&-e\int_P\Tr\ls\G^i(P+K,P)S(P)\g^j S(P+K)\rs,\\
G_{AV}^{ij}&=&-e\int_P\Tr\ls\G^i(P+K,P)S(P)\g^j\g_5 S(P+K)\rs. \quad
\end{eqnarray}
The effective vertex in the above is given by \fig{diag:vertex},
\begin{eqnarray}
\label{vertexeq}
&&\G^\m(P+K,P)=\g^\m+e^2\int_Q\g^\r S(P+K+Q)\non
&&\times\G^\m(P+K+Q,P+Q)S(P+Q)\g^\s {^* D_{\r\s}}(Q). \quad
\end{eqnarray}
The $^*D_{\r\s}(Q)$ is the HTL propagator for the photon, while $S(P)$
is the electron propagator at nonzero $\m_V$ and $\m_A$,
\begin{eqnarray}
S(P)&=&\frac{-1}{\g_0(p_0+\m_V+\m_A\g_5)-{\bm \g}\cdot\bp-\g^\m\S_\m(P)}\non
&=&\sum_{s=\pm}\Xi_s\frac{-\g\cdot(P_s-\S_s)}{(P_s-\S_s)^2}, \quad
\end{eqnarray}
where $\Xi_\pm=(1\pm\g_5)/2$ is the chirality projection,
$P_\pm^\m=(P_\pm^0, \bp)$ with $P_\pm^0=p_0+\m_\pm$ and $\m_\pm=\mu_V\pm\m_A$.
Substituting all these into the Kubo formulas and following
essentially the steps as in Refs.~\cite{ValleBasagoiti:2002ir,Aarts:2002tn}, we finally obtain
\begin{eqnarray}
\s&=&\sum_{s,a=\pm}\s_{sa},\\
\s_e&=&\sum_{s,a=\pm}s\s_{sa},\\
\s_{sa}&=&-a\frac{e}{3}\int\frac{d^3\bp}{(2\p)^3}n_F'(p-a\m_s)\c_a^s(p),
\end{eqnarray}
with $n_F(x)=1/[\exp{(x/T)}+1]$ and $\c_a^s(p)={\cal D}_a^s(\bp)/\G_a^s$. The ${\cal D}_a^s$ is  defined through
$2p^i{\cal D}_+^s(\bp)=\bar{u}_s(\bp)\G^i_+(\bp)u_s(\bp),\; 2p^i{\cal D}_-^s(\bp)=\bar{v}_{-s}(-\bp)\G^i_-(\bp)v_{-s}(-\bp)$,  and
$\G_{\pm}^s=-2\im\S^R_{\pm s}(P^0_s= \pm E_p,\bp)$ is the decay width for fermions ($+$) and
antifermions ($-$) of chirality $s$.

The vertex integral equation (\ref{vertexeq}) can be finally recast into a differential equation for $\c_a^s(p)$,
\begin{eqnarray}
1&=&\frac{2\a m_s^2\ln(1/e)}{p}[n_B(p)+n_F(\m_s)]\c_a^s(p)\non&&+
\frac{\a Tm_D^2\ln(1/e)}{p^2}\bigg[\c_a^s(p)-\frac{p^2}{2}\frac{d^2\c_a^s(p)}{dp^2}\non&&-\lb1-\frac{p}{2T}(1-2n_F[p-a\m_s])\rb p\frac{d\c_a^s(p)}{dp}\bigg],
\end{eqnarray}
where $n_B(x)=1/[\exp{(x/T)}-1]$ is the Bose-Einstein function, $\a=e^2/4\p$, $m_s^2=\frac{e^2}{8}\lb T^2+\frac{\m_s^2}{\p^2}\rb$
is the effective mass of electrons, and $m_D^2=\frac{e^2}{6}\sum_{s=\pm}\lb T^2+\frac{3\m_s^2}{\p^2}\rb$ is the Debye screening mass
of photons.
We note that the leading-order terms in the effective vertex and those in the fermion decay width cancel
each other and as a result the leading-log results for conductivities are actually at the order $\hat{O}(1/[e^3\ln(1/e)])$ rather than $\hat{O}(1/[e\ln(1/e)])$ as expected from $\s_{sa}\sim e/\G^s_a$.

Following Ref.~\cite{Arnold:2000dr}, we solve this differential equation numerically by using the variational method. This variational scheme converges very fast, and at very high precision we obtain (up to $\hat{O}(\m_s/T)^2$ order)
\begin{eqnarray}
&& \s_{sa} \approx \frac{T}{e^3\ln(1/e)}\bigg(3.9238+ 4.59867 a\frac{\m_s}{T}+ 2.56237 \frac{\m_s^2}{T^2} \nonumber \\
 && \qquad   -0.62224\frac{\m_{V}^2+\m_{A}^2}{T^2}\bigg),
\end{eqnarray}
and thus
\begin{eqnarray}
\label{sigma}
\s&\approx&\frac{T}{e^3\ln(1/e)}\lb15.6952+7.76052\frac{\m_V^2+\m_A^2}{T^2}\rb,\\
\label{sigma_e}
\s_e&\approx&20.499\frac{\m_V\m_A}{T^2}\frac{T}{e^3\ln(1/e)}.
\end{eqnarray}
In the limit $\m_s\to 0$, our
result for $\s$ is in agreement with $\s\approx15.6964\frac{T}{e^3\ln(1/e)}$ obtained in Ref.~\cite{Arnold:2000dr}.

{\em Coupled evolution of the two currents.---} As seen in Eq.(\ref{vaeb}), with the presence of external electromagnetic fields, the vector and axial currents mutually induce each other and get entangled in a nontrivial way. It is of great interest to understand the coupled evolution of small fluctuations of the two currents. A very good example is the aforementioned CMW \cite{Kharzeev:2010gd,Burnier:2011bf} in which the fluctuations of vector and axial currents are coupled together by external $\vec B$ field to form a propagating wave. Now the new CESE effect introduces nonlinearity (through the $\mu_V\mu_A$ term) and makes the problem more nontrivial.

Let us consider a thermal QED or QCD plasma in the static and homogeneous external $\vec E, \vec B$ fields and study the coupled evolution of the small fluctuations in vector and axial charge densities. The presence of vector
density and current will induce additional electromagnetic fields so that $\vec E_{\rm tot}=\vec E +\d\vec E$ and $\vec B_{\rm tot}=\vec B +\d\vec B$ with $\d \vec E\propto ej_V^0$ and $\d \vec B\propto e \vec j_V$. As in the case of CMW, one can first replace the chemical potentials in Eq.(\ref{vaeb}) with the corresponding charge densities, $\mu_{V,A}=\alpha_{V,A} j_{V,A}^0$, where the  $\alpha_{V,A}$ are the susceptibilities defined as $\alpha_{V,A} \equiv \partial \mu_{V,A} / \partial j_{V,A}^0$. These relations are valid as long as the chemical potentials are small compared with temperature $T$. Then, combining Eq.(\ref{vaeb}) with the currents' continuity equations $\partial_t j_{V,A}^0 + \vec \bigtriangledown \cdot \vec j_{V,A}=0$ and Maxwell's equation $\vec \bigtriangledown\cdot\vec E_{\rm tot}=ej_V^0$ and $\vec \bigtriangledown\cdot\vec B_{\rm tot}=0$, one can obtain
\begin{eqnarray}
&&\partial_t j_V^0 + e\sigma_0 j_V^0
+ \sigma_5 \alpha_A (\vec B \cdot \vec \bigtriangledown ) j_A^0 \nonumber \\
&&\quad\!\!\!\!\! +2\sigma_2\alpha_V^2 j_V^0(\vec E \cdot\vec\bigtriangledown )j_V^0+2\sigma_2\alpha_A^2 j_A^0(\vec E \cdot\vec\bigtriangledown) j_A^0=0\, , \nonumber \\
   &&\partial_t j_A^0
+\sigma_5 \alpha_V (\vec B \cdot \vec \bigtriangledown ) j_V^0 +  \chi_e \alpha_V \alpha_A  j_V^0( \vec E \cdot \vec \bigtriangledown )  j_A^0  \nonumber \\
&&\quad +  \chi_e \alpha_V \alpha_A  j_A^0( \vec E \cdot \vec \bigtriangledown )  j_V^0
 =0     \, . \quad
\end{eqnarray}
where we have introduced $\sigma_0$ and $\sigma_2$ defined through the small chemical potential expansion of $\sigma$,
$\sigma=\sigma_0+\sigma_2(\mu_V^2+\mu_A^2)$. In arriving at the above, we have made the following approximations: first, we keep only terms up to $\hat{O}(j_{V,A}^2)$ order (such that terms like $j^0 \delta \vec E \cdot \bigtriangledown j^0 \sim \hat{O}(j^3)$ and $\sigma_2 \mu^2  j_V^0$ are dropped); second, we assume external fields are extremely strong such that the magnetic field feedback terms $\sim \sigma_5 \alpha \delta \vec B \cdot \bigtriangledown j^0 \sim \sigma_5\alpha e (j \partial j) $ are negligible to other $\hat{O}(j^2)$ terms $\sim\chi_e \alpha^2 E (j\partial j)$ and $\sim\sigma_2 \alpha^2 E (j\partial j)$. We will next linearize the above equations by considering fluctuations on top of small uniform background vector and axial densities $n_V$ and $n_A$, so that $j_{V,A}^0=n_{V,A}+\d j_{V, A}^0$ with $\d j_{V,A}^0\ll n_{V,A}$ being fluctuations. Note that $n_{V,A}$ themselves must still be small (compared to $T^3$) to ensure  the linear relations between $j^0_{V,A}$ and $\mu_{V,A}$. Strictly speaking, only a uniform axial density $n_A$ is a static solution of the above equations, while a uniform vector density $n_V$ suffers from the damping term $e\sigma_0 j_V^0$ and is only approximately static on time scale shorter compared with $1/e\sigma_0$. Keeping only linear terms in $\d j_{V,A}^0$, we obtain
\begin{eqnarray}
\label{vacoupled}
&&\partial_t \d j_V^0 + e\sigma_0 \d j_V^0
+ \sigma_5 \alpha_A (\vec B\cdot \vec \bigtriangledown ) \d j_A^0 \nonumber \\
&&\quad +2\sigma_2\alpha_V^2 n_V(\vec E\cdot\vec\bigtriangledown ) \d j_V^0+2\sigma_2\alpha_A^2 n_A(\vec E\cdot\vec\bigtriangledown) \d j_A^0=0\, , \nonumber \\
   &&\partial_t \d j_A^0
+\sigma_5 \alpha_V (\vec B\cdot \vec \bigtriangledown ) \d j_V^0 +  \chi_e \alpha_V \alpha_A  n_V( \vec E\cdot \vec \bigtriangledown )  \d j_A^0  \nonumber \\
&&\quad +  \chi_e \alpha_V \alpha_A  n_A( \vec E \cdot \vec \bigtriangledown )  \d j_V^0
 =0     \, . \quad
\end{eqnarray}

In order to find possible normal modes, we make the Fourier transformation of these equations. Using $\d j_{V,A}^0=\int_{\omega,\vec k} C_{V,A}(\omega,\vec k)\, e^{-i(\omega t - \vec k\cdot \vec x)}$, we obtain from Eqs. (\ref{vacoupled}) the following relations:
\begin{eqnarray}
\label{vamodes}
&&\omega C_V+ ie\sigma_0 C_V
- \sigma_5 \alpha_A (\vec B\cdot \vec k ) C_A \nonumber \\
&&\quad -2\sigma_2\alpha_V^2 n_V(\vec E\cdot \vec k) C_V -2\sigma_2\alpha_A^2 n_A(\vec E\cdot \vec k) C_A=0\, , \nonumber \\
&&\omega C_A
-\sigma_5 \alpha_V (\vec B\cdot \vec k ) C_V- \chi_e \alpha_V \alpha_A  n_A(\vec E\cdot \vec k)  C_V  \nonumber \\
&&\quad - \chi_e \alpha_V \alpha_A  n_V(\vec E\cdot \vec k)  C_A
 =0   \, . \quad
\end{eqnarray}
Without loss of generality, we can always assume $\vec B$ is along the $z$-axis, i.e., $\vec B = B \hat{\bf z}$ while $\vec E =  E \hat{\bf e}$. The dispersion relation obtained from Eq. (\ref{vamodes}) can be expressed as
\begin{eqnarray}
\label{dispersion}
&&\o=-\frac{1}{2}\ls ie\s_0-v_+(\hat{\bf e}\cdot\vec k)\rs\non&&\quad\pm
\frac{1}{2}\sqrt{\ls ie\s_0-v_-(\hat{\bf e}\cdot\vec k)\rs^2+4{\cal A}_\c(\vec k)},
\end{eqnarray}
where $v_\pm=v_v\pm v_a$ with $v_v=2\sigma_2\alpha_V^2 n_V E$ and $v_a = \chi_e \alpha_V  \alpha_A n_V E$, and
\begin{eqnarray}
{\cal A}_\c(\vec k)&=&\ls\s_5\a_A B(\hat{\bf z}\cdot\vec k)+2\s_2\a_A^2 n_A E(\hat{\bf e}\cdot\vec k)\rs\non&&\!\!\!\!\!\times\ls\s_5\a_V B(\hat{\bf z}\cdot\vec k)+\c_e\a_V\a_A n_A E(\hat{\bf e}\cdot\vec k)\rs.
\end{eqnarray}
To manifest the physical meaning of the solutions in (\ref{dispersion}), let us consider the following two special cases:

(1) The case with only $\vec B = B \hat {\bf z}$ and $\vec E={\bf 0}$.\\ Equation (\ref{dispersion}) reduces to
$\omega = \pm \sqrt{(v_\chi k_z)^2 - (e\sigma_0/2)^2 } - i (e\sigma_0/2)$ with speed $v_\chi = \sigma_5 \sqrt{\alpha_V \alpha_A} B$. When $v_\chi k_z \gg e\sigma_0/2$ we get two well-defined propagating modes $\omega\approx  \pm v_\chi k_z -  i (e\sigma_0/2)$. These are generalized CMWs which reduce to the CMW in \cite{Kharzeev:2010gd} when $\s_0=0$ and $\alpha_V=\alpha_A$. When $v_\chi k_z \le e\sigma_0/2$ the two modes become purely damped.

(2) The case with only $\vec E =E \hat {\bf z}$ and $\vec B={\bf 0}$. \\
First, we consider a background without vector density, i.e., $n_V=0$. In this case, we find two modes from (\ref{dispersion}),
\begin{eqnarray}
\omega = \pm \sqrt{(v_e k_z)^2 - (e\sigma_0/2)^2 } - i (e\sigma_0/2)
\end{eqnarray}
with $v_e = \a_A n_A \sqrt{2\s_2\c_e\alpha_V \alpha_A} E$. Similar to the CMWs, when $v_e k_z \gg e\sigma_0/2$ there are two  well-defined modes $\omega\approx  \pm v_e k_z -  i (e\sigma_0/2)$ from CESE that propagate along $\vec E$ field and can be called the chiral electric waves (CEWs).  They become damped when $v_e k_z \le e\sigma_0/2$. \\
Second, if the background contains no axial density, i.e., $n_A=0$, then we see that the vector and axial modes become decoupled, and Eq. (\ref{dispersion}) leads to
\begin{eqnarray}
\o_V(\vec k)&=& v_v k_z - i (e\s_0) \; ,\non
\o_A(\vec k)&=& v_a k_z.
\end{eqnarray}
The first solution $\o_V(\vec k)$  represents a ``vector density wave" (VDW) with speed $v_v=2\sigma_2\alpha_V^2 n_V E$ that transports vector charges along $\vec E$ field but will be damped on a time scale $\sim 1/(e\s_0)$. The second solution $\o_A(\vec k)$ is a new  mode arising from CESE and represents a propagating ``axial density wave" (ADW) along $\vec E$ with speed $v_a = \chi_e \alpha_V  \alpha_A n_V E$ and without damping.

{\em Summary and discussions.---}In summary, we have found a new mechanism for the generation of axial current by external electric field in a conducting matter with nonzero vector and axial charge densities, which we call the chiral electric separation effect. We have computed the CESE conductivity coefficient in a QED plasma and also studied possible collective modes arising from it.

\begin{figure}[!htb]
\begin{center}
\includegraphics[width=8cm]{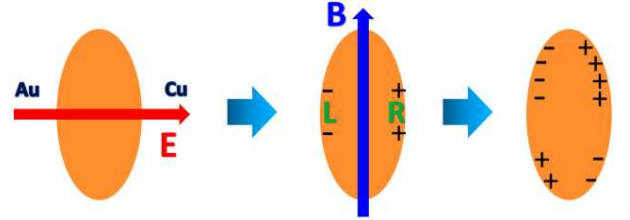}
\caption{A schematic illustration for CESE-induced net charge distribution and  correlation patterns in Cu + Au collision.
\vspace{-0.25in}
} \label{diag:ca}
\end{center}
\end{figure}


We end by discussing possible observable effects induced by CESE in heavy ion collisions. In the created hot QGP there can be both vector and axial charge densities from fluctuations and topological transitions. There are also very strong electric fields during the early moments of heavy ion collisions ~\cite{Voronyuk:2011jd,Bzdak:2011yy,Deng:2012pc,Bloczynski:2012en}. One particularly interesting situation is in the Cu + Au collisions (see Fig. \ref{diag:ca}), where due to the asymmetric nuclei (rather than from fluctuations) there will be a strong $\vec E$ field directing from the Au nucleus
to the Cu nucleus ~\cite{Hirono:2012rt}. In this case the $\vec E$ field will lead to both an in-plane charge separation via (\ref{con}) and an in-plane chirality separation via (\ref{cese}). The resulting in-plane axial dipole will then further  separate charges via CME along the magnetic field in the out-of-plane direction, and cause an approximate quadrupole at certain angle $\Psi_q$ in between in- and out-of-plane . We therefore expect a highly nontrivial charge azimuthal distribution pattern $ \delta N_{\pm} (\phi) \sim d \cos(\phi) + q \cos(2\phi-2\Psi_q)$ with the dipole term due to usual conductivity and the quadrupole term due to CESE and CME effects. This pattern may possibly be measured either via charged pair correlations or the charged multiple analysis \cite{Bloczynski:2012en}. Quantitative predictions will require proper modeling of the QGP and solving the \eq{vacoupled}, which will be reported in a future work.

\vskip0.2cm

{\em Acknowledgments---}
We thank Dmitri Kharzeev, Shu Lin, Misha Stephanov, and Ho-Ung Yee for helpful discussions.
J.L. is grateful to the RIKEN BNL Research Center for partial support.

 \vfil

\end{document}